# Local Electrical Stress-Induced Doping and Formation of 2D Monolayer Graphene P-N Junction

Tianhua Yu[†], Chen-Wei Liang[†], Changdong Kim[†], and Bin Yu[*,†]

[†] College of Nanoscale Science and Engineering, State University of New York, Albany, NY 12203

[*] Corresponding Author.  byu@uamail.albany.edu

**ABSTRACT**  We demonstrated doping in 2D monolayer graphene via local electrical stressing. The doping, confirmed by the resistance-voltage transfer characteristics of the graphene system, is observed to continuously tunable from N-type to P-type as the electrical stressing level (voltage) increases. Two major physical mechanisms are proposed to interpret the observed phenomena: modifications of surface chemistry for N-type doping (at low-level stressing) and thermally-activated charge transfer from graphene to $SiO_2$ substrate for P-type doping (at high-level stressing). The formation of P-N junction on 2D graphene monolayer is demonstrated with complementary doping based on locally applied electrical stressing.

Graphene, a 2D allotrope of carbon, exhibits immense potential in nanoscale device applications because of its extraordinary properties including ultra-high carrier mobility[1] and thermal conductivity,[2] chemical inertness,[3] and electromechanical robustness.[4] In addition, its two-dimensional (2-D) planar nature is ideal to implement "super-thin-body" transistor, suppressing the so-called short-channel effects that impede the scaling of semiconductor-based transistors.[5] The electric field effect in graphene has its root in the unique energy band structure derived from the 2-D honeycomb carbon lattice. Carrier transport is described by the Dirac equation at low energies.[6] The $E$-$k$ dispersion of monolayer graphene displays a linear and conical feature. The conduction and valence bands are separated by a charge neutrality point (CNP, also called the Dirac point) in which the available number of electronic states is zero.[7] Therefore, graphene is a zero-bandgap semimetal and can be either electron-conducting ($E_f > E_{CNP}$) or hole-conducting ($E_f < E_{CNP}$), depending on the Fermi level ($E_f$) position with respect to the CNP energy ($E_{CNP}$).

Complementary doping in graphene is demanded in the implementation of ultra-low-power logic and emerging novel devices based on P-N junction structure, such as tunneling FET and Veselago lens, yet it still remains as one of the major challenges.[8] A number of doping approaches have been proposed for graphene. Chemical methods are used to produce large-area doping with other elements, e.g., oxygen, nitrogen or hydrogen.[9] However, chemical doping compromises carrier mobility as a result of exacerbated charge scattering at increased impurity level.[10] Recently, doping effect in graphene under high-level electrical stress was reported and substantial shift of the CNP was observed.[11] The electrically induced doping method points to the possible route to achieve on-chip, site-specific doping without complicated material process and device performance degradation. However, the reported results are limited to one-type-only doping.[12,13].

We conduct experimental study with the goal to achieve complementary type of doping in graphene. The device structure used is schematically shown in Fig. 1a. Graphene flakes were obtained by micromechanical exfoliation onto degenerately P-type doped Si substrate with a thermal $SiO_2$ layer (100nm thick) on the surface. Monolayer graphene was identified by optical microscopy and further confirmed by micro-Raman spectroscopy.[14] The graphene sheet is contacted by electrodes using metal (Au) deposition, electron-beam lithography patterning, and subsequent lift-off process. The devices is a three-terminal FET structure with two electrodes as source (S) and drain (D) and the doped Si substrate as the gate (G) electrode. Devices were thermally annealed in $H_2$ / Ar (15 sccm / 200 sccm) at 200°C for two hours prior to the electrical characterization in order to improve the metal / graphene contact quality.[15] All electrical measurements were performed at room temperature in vacuum (~ $10^{-6}$ Torr).

Fig. 1b shows the measured transfer characteristics (graphene channel resistance vs. gate voltage, or $R$ - $V_G$) of a back-gated graphene FET before and after electrical stressing. The graphene channel length and width are 6 µm and 3 µm, respectively. Prior to electrical stressing, the device showed a P-type conduction with CNP found at $V_G$ = 12 V (curve ①). The contact resistance is around 11 kΩ, assuming graphene channel resistance is negligible (as compared with contact resistance) at high back-gate voltage. The P-type conduction is typically observed in the as-fabricated graphene FETs and commonly attributed to moisture or other ambient contamination. To add electrical stressing, we performed a drain voltage sweeping from $V_{DS}$ = 0 V to $V_{DS}$ = 10 V (at $V_G$ = 0 V). The maximum current resulting from the voltage sweeping was 1.2 mA. Multiple changes in the transfer characteristic were observed after the electrical stressing (curve ②). First, the CNP was shifted to $V_G$ = -9 V. Second, the contact resistance was dramatically reduced to 4.5 kΩ. Third, the carrier mobility derived from the transfer curve was found to increase from 700 to 1300 $cm^2V^{-1}s^{-1}$. The negative shift of the CNP represents a transition from hole-dominant conduction to electron-dominant conduction after the electrical stressing. The device was further stressed with higher D/S voltage sweeping and larger current (15 V / 2.2 mA). A double-peak feature was displayed in the transfer characteristic after the stressing (curve ③): The resistance peak was shifted to $V_G$ = -16.7 V and a new peak at $V_G$ = 18.7V was also observed. The double-peak phenomenon shown in ③ indicates that a P-N junction within the graphene channel was formed. Interestingly, further stressing steps resulted in shifting of the CNPs towards positive voltages (curves ④ and ⑤). After electrical stressing at 20 V / 3.9 mA, two CNPs (at $V_G$ = 0 and $V_G$ = 23 V) were observed in a more balanced manner (Curve ④), both showing positive shifting. After stressing the device at 25 V / 5.2 mA, only one CNP at $V_G$ = 28 V now remained (curve ⑤), showing a transition from a P-N junction to a fully P-doped graphene channel throughout the channel. It should be noted that the contact resistance continued to decrease with increasing level of electrical stressing.

The changes in the transfer characteristics as shown in Fig. 1b represent multiple processes during the electrical stressing. The graphene channel was firstly converted from P-type to N-type (①→②), followed by the formation of P-N junction and the eventual switch back to P-type (②→⑤). Detailed analyses of the data reveals that the observed complementary doping is actually a two-step process: The mechanism of the first conversion (①→②, from P-type to N-type) is attributed to the modification of surface chemistry by removing residuals absorbed on graphene surface (e.g., $HO_2$ or other ambient gas molecules, as shown in Fig. 2a).[16] The shift of one of the CNP in ③ from $V_G$ = -9 V to -16.7 V can be explained with the removal of such residuals due to Joule heating induced by the electrical stressing. As the electrical stressing level increases (③→④→⑤), the second process kicks in. The high vertical electric field in the drain side would cause electrons to gain enough energy to be injected from graphene into deep-level trap states in $SiO_2$ near the graphene/substrate interface, making graphene more towards P-type doped as was also observed previously.[11] When the P-type doping level is relatively low, a P-N junction is formed with P-type region is located on the drain side. At the highest level of stressing in our experiment (case ⑤), hole concentration in graphene is high enough to make the whole channel purely

P-type doped. In addition to the first-time demonstration of complementary type of doping in graphene via electrical stressing technique, our results also show the possibility to form P-N junction in a reasonably controlled way. After repeating the same experiment for more 50 graphene samples, we observed the same trend of continuous change of doping configuration, suggesting the great potential of a tunable 2D carbon material system. We have also noticed that further increases of stressing level eventually lead to permanent failure, i.e., graphene flakes were found to be physically broken.[17] The thermal-activated charge transfer from graphene to $SiO_2$ (Fig. 2b) could share the similar mechanism as in the reported doping effect of thermally annealed graphene.[18] While it is beyond the scope of this work, in-depth study is needed to provide microscopic picture of the charge transfer between graphene and $SiO_2$.

With the electrical doping technique, we further demonstrate graphene-based P-N junction created by electrical stressing. The test device consists of a sheet of graphene monolayer and six contacting electrodes, denoted as A, A', B, B', C and C' (Fig. 3a). The A-A' and B-B' electrode pairs are designed to have a narrow spacing (< 1 μm, as shown in Fig. 3b) between the electrodes in each pair to enhance the doping effect via electrical stressing. An N-type doped region was created by applying low-level stressing (10 V) across A and A', while a P-type doped region was induced by high-level stressing (30 V) across B and B'. These doping processes were confirmed by measuring the corresponding transfer characteristics. As shown in Fig. 3c, the post-stressing CNP of the device using A/A' as the S/D electrodes is at $V_G$ = -17V, confirming the local N-type doped region between the A-A' electrode pair, while the post-stressing CNP of the B-B' device is at $V_G$ = 4.6V, confirming the local P-type doping. No double-peak feature was observed in these curves, showing each region is single-type doped. Next, when measuring the transfer characteristic across the C-C' electrode pair, a resistance peak at $V_G$ ~ -14 V with a broad shoulder at $V_G$ ~ 4V was observed (Fig. 3d). This transfer curve is the superposition of two Lorentzian-type curves after performing numerical fitting. The peak positions of the two fitting curves are found to be at $V_G = -17.5V$ and $V_G = 4.7V$, respectively, close to the positions of the CNPs observed in the individual transfer characteristic of A-A' and B-B' pairs (Fig. 3c). This consistency confirms the formation of P-N junction induced by local electrical doping via A-A' and B-B' pairs, as reflected in the characteristics measured in C-C' pair.

To investigate the doping impact on electrical transport property in graphene, carrier mobility is extracted from the fabricated back-gated FETs. Fig. 4 shows the measured mobility as a function of the stressing voltage. Mobility is calculated using the following relation: $\mu = 1/(n\rho e)$, where $n$ is the carrier density, $\rho$ is the resistivity, and $e$ is the electronic charge. We observe that carrier mobility is improved as a result of low-level voltage stressing. The improvement of mobility is attributed to the reduction of impurity-induced scattering, as confirmed by the absence of the D-band in the Raman spectra. The fact suggests that electrical stressing can be used to provide rapid doping in graphene with minimal degradation in carrier transport. However, at high-level of electrical stressing the carrier mobility is degraded, possibly due to the short-range scattering caused by deep-level trapped charges near the surface of $SiO_2$ substrate.

In summary, we have demonstrated complementary doping in 2D monolayer graphene by local electrical stressing. The possible mechanisms of doping resulting from electrical stressing are surface chemistry modification (at low-level stressing) for N-type doing and thermal-activated charge transfer (at high-level stressing) for P-type doping. With a specially designed test structure we successfully demonstrated the formation of graphene P-N junction using locally applied electrical stressing. Our results suggest that complimentary type of doping can be achieved by employing the illustrated two-step process. The experiment unveils the potential of on-site local doping manipulation or engineering for implementing next-generation of 2D nano-carbon electronics.

**Acknowledgment.** The research was partially supported by National Science Foundation (NSF) grants (ECCS-1002228 and ECCS-1028267) and IBM Faculty Award.

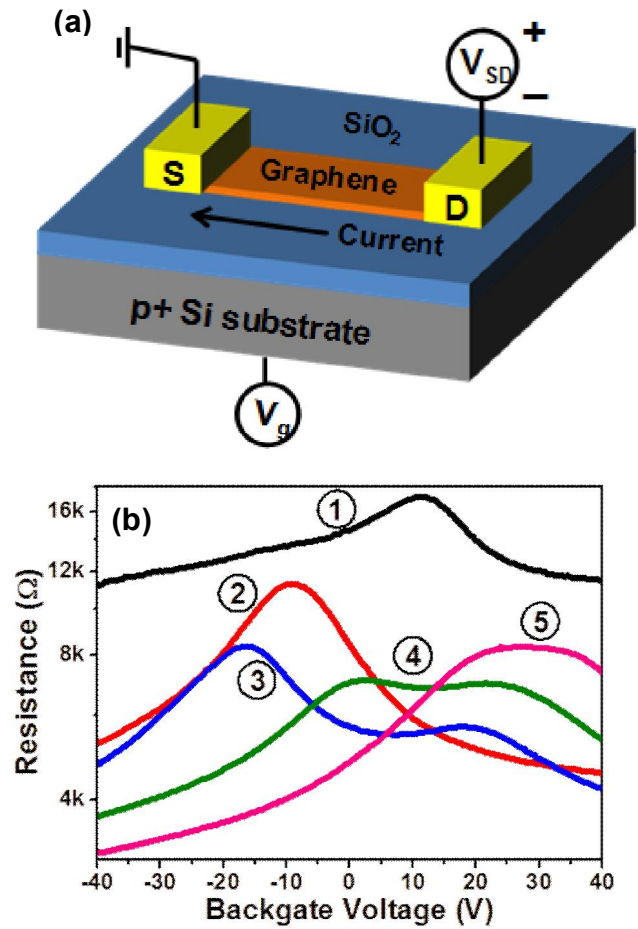

**FIG. 1: (a)** Schematic structure of graphene FET on $SiO_2$ substrate with heavily p-doped Si substrate serving as the back gate. Here *S* and *D* refer to the source and drain metal contacts, respectively. **(b)** Measured transfer characteristic ($R$ vs. $V_G$ curve) after applying DC stressing voltage sweeping from 0 V to the following maximum value: 0 V ("1"), 10 V ("2"), 15 V ("3"), 20 V ("4"), and 25 V ("5").

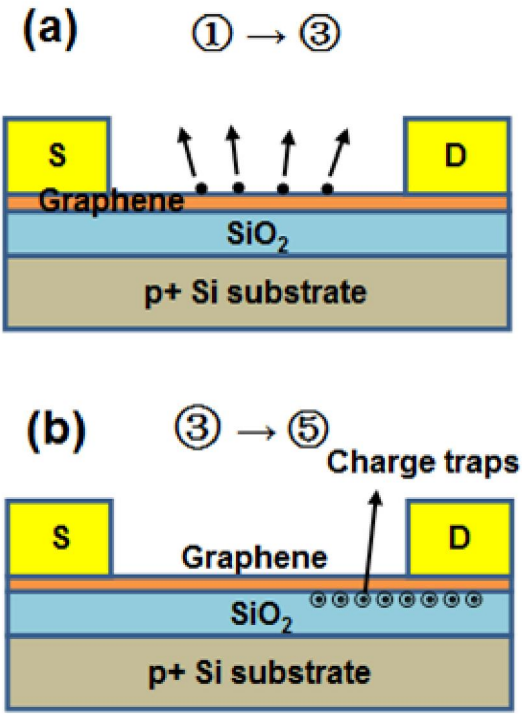

**FIG. 2:** Schematic illustration of the two mechanisms of graphene doping with electrical stressing. **(a)** Removal of absorbed charge impurities on graphene surface (at low-level stressing). **(b)** Thermally activated charge transfer from graphene to $SiO_2$ substrate (at high-level stressing).

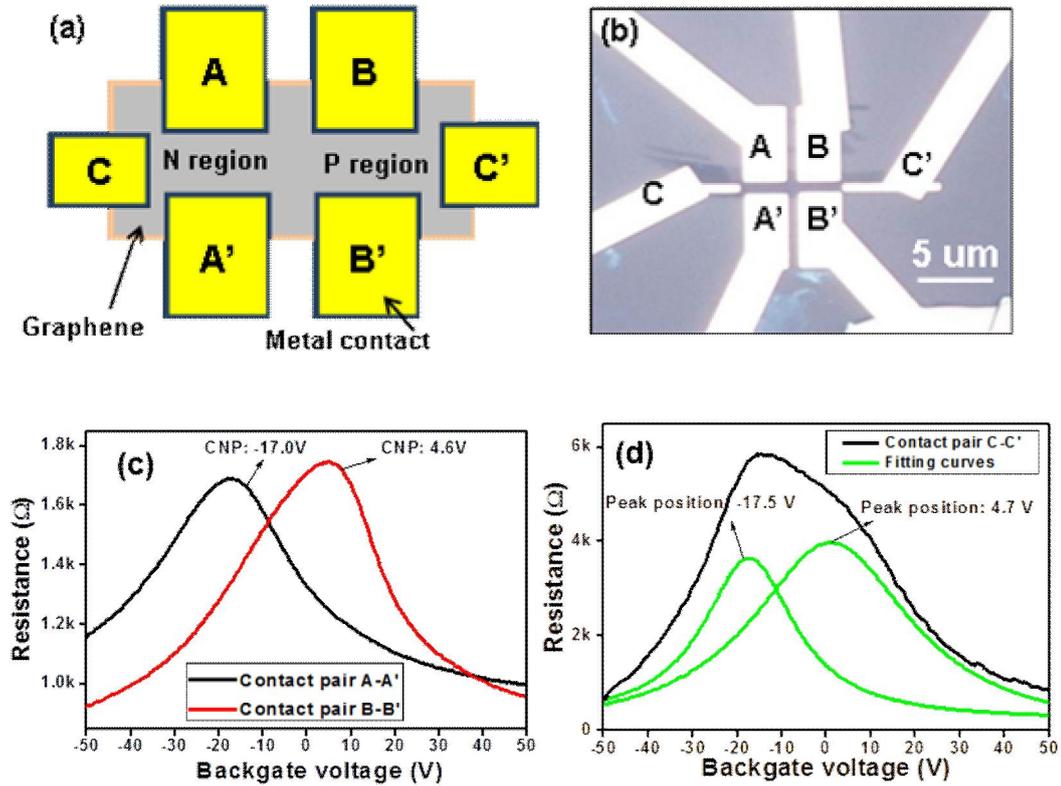

**FIG. 3:** Formation of P-N junction using local electrical stressing induced doping. **(a)** Schematic of the testing structure (top-view layout) with 3 pairs of contacting electrodes (6 terminals). **(b)** Optical image of the fabricated graphene device. **(c)** R - $V_G$ transfer characteristics measured from paired structures A-A' and B-B', respectively. **(d)** R - $V_G$ transfer characteristic measured from the C-C' paired structure showing double-peak feature as indicated by the fitted curves.

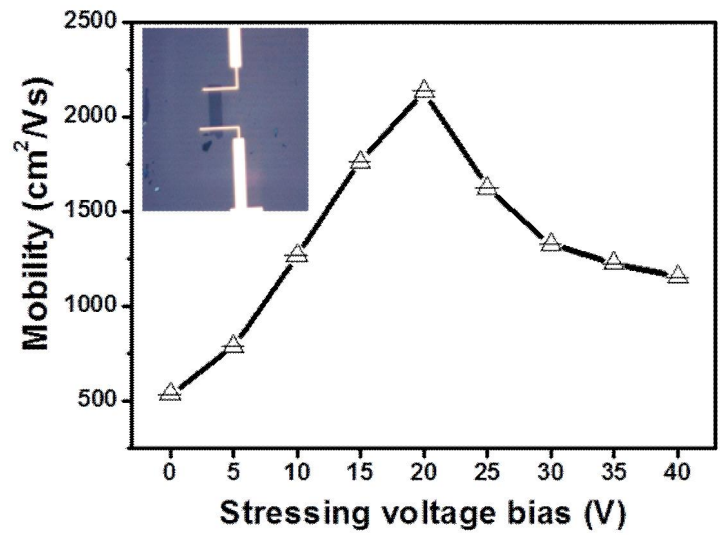

**Fig. 4:** Measured carrier mobility in monolayer graphene as a function of stressing voltage. The mobility is extracted from the R - $V_G$ transfer characteristics of the back-gated graphene FET. Inset is the optical image of the actual device.